\documentclass[twocolumn]{article}

\usepackage{amsmath, amssymb}   
\usepackage{graphicx}           
\usepackage{geometry}           
\usepackage{hyperref}           
\usepackage{titling}            
\usepackage{indentfirst}
\usepackage{rotating}
\usepackage{multicol}  

\usepackage[backend=biber, style=ieee]{biblatex} 
\pretitle{
    \begin{center}
    \rule{\textwidth}{1.5pt} \\[0.5ex] 
    \LARGE \bfseries 
}
\posttitle{
    \\[0.5ex]
    \rule{\textwidth}{1.5pt} \\  
    \vspace{1em}
    \end{center}
}

\title{BreachSeek: A Multi-Agent Automated Penetration Tester}

\author{
    \textbf{Ibrahim AlShehri}\textsuperscript{ 1 *} \\
    \small\texttt{ibrahimalshehri@pm.me} \\
    \and
    \textbf{Adnan AlShehri}\textsuperscript{ 1 *} \\
    \small\texttt{adnan66b@gmail.com} \\
    \and
    \textbf{Abdulrahman AlMalki}\textsuperscript{ 1 *} \\
    \small\texttt{almalki\_abdulrahman@outlook.com} \\
    \and
    \textbf{Majed Bamardouf}\textsuperscript{ 1 *} \\
    \small\texttt{majedTB12@gmail.com} \\
    \and
    \textbf{Alaqsa Akbar}\textsuperscript{ 1 *} \\
    \small\texttt{alaqsaakbar@hotmail.com} \\
}

\date{\normalsize\textsuperscript{1} King Fahd University of Petroleum and Minerals (KFUPM) \\
\textsuperscript{*} Equal contribution}

\begin{document}

\maketitle

\begin{abstract}
\textit{
The increasing complexity and scale of modern digital environments have exposed significant gaps in traditional cybersecurity penetration testing methods, which are often time-consuming, labor-intensive, and unable to rapidly adapt to emerging threats. There is a critical need for an automated solution that can efficiently identify and exploit vulnerabilities across diverse systems without extensive human intervention. BreachSeek addresses this challenge by providing an AI-driven multi-agent software platform that leverages Large Language Models (LLMs) integrated through LangChain and LangGraph in Python. This system enables autonomous agents to conduct thorough penetration testing by identifying vulnerabilities, simulating a variety of cyberattacks, executing exploits, and generating comprehensive security reports. In preliminary evaluations, BreachSeek successfully exploited vulnerabilities in exploitable machines within local networks, demonstrating its practical effectiveness. Future developments aim to expand its capabilities, positioning it as an indispensable tool for cybersecurity professionals.
}
\end{abstract}

\section{Introduction}
The rapid evolution of cyber threats has underscored the limitations of traditional cybersecurity practices, particularly in the domain of penetration testing. Manual penetration testing, while thorough, is inherently time-consuming and increasingly ineffective in keeping pace with the growing sophistication and diversity of cyberattacks. In an era where networks and applications are constantly exposed to new vulnerabilities, there is a pressing need for automated solutions that can efficiently identify, exploit, and report on these weaknesses.

Recent advancements in Artificial Intelligence (AI) and Natural Language Processing (NLP) have opened up new possibilities for automating complex tasks, including cybersecurity. Large Language Models (LLMs), renowned for their capabilities in natural language understanding and generation, have demonstrated the potential to perform tasks that traditionally required significant human expertise. Despite these advancements, the application of LLMs in cybersecurity, particularly for automating penetration testing, remains largely underexplored, presenting an opportunity to revolutionize how security assessments are conducted.

BreachSeek addresses this critical gap by introducing an AI-driven, multi-agent software platform specifically designed to automate penetration testing for websites and networks. The platform leverages the power of LLMs through LangChain and LangGraph in Python, allowing autonomous agents to identify vulnerabilities, simulate a variety of sophisticated cyberattacks, and execute exploits with minimal human intervention. By automating these processes, BreachSeek not only accelerates the penetration testing workflow but also enhances the accuracy and comprehensiveness of the results, providing a robust solution to the ever-evolving landscape of cybersecurity threats.

One of the key technical innovations in BreachSeek is the use of multiple AI agents, each with a distinct focus, to manage the complexity and breadth of tasks involved in penetration testing. This approach ensures that the system avoids running out of context window, a common limitation in LLMs, and allows for the separation of concerns. Each agent is tasked with a specific aspect of the testing process, ensuring a high level of specialization and accuracy. This design principle not only optimizes the performance of individual agents but also contributes to the overall efficiency and effectiveness of the platform.

The platform's scalability further enhances its utility, enabling it to be deployed in a wide range of environments, from small to large-scale networks. By deploying multiple agents in different containers, BreachSeek can efficiently manage large volumes of data and complex network architectures, making it adaptable to various cybersecurity needs. This scalability is particularly beneficial for organizations that operate in sectors with high security demands, such as finance, healthcare, and government, where the ability to rapidly and accurately identify vulnerabilities is crucial.

In summary, BreachSeek represents a significant advancement in the field of automated cybersecurity penetration testing. By combining the power of AI-driven agents with the flexibility and scalability required in modern network environments, BreachSeek offers a comprehensive solution that addresses the limitations of traditional penetration testing methods. As cyber threats continue to evolve, tools like BreachSeek will become increasingly vital in ensuring the security and resilience of digital infrastructure.

\begin{figure*}[ht]
    \centering
    \includegraphics[width=1\linewidth]{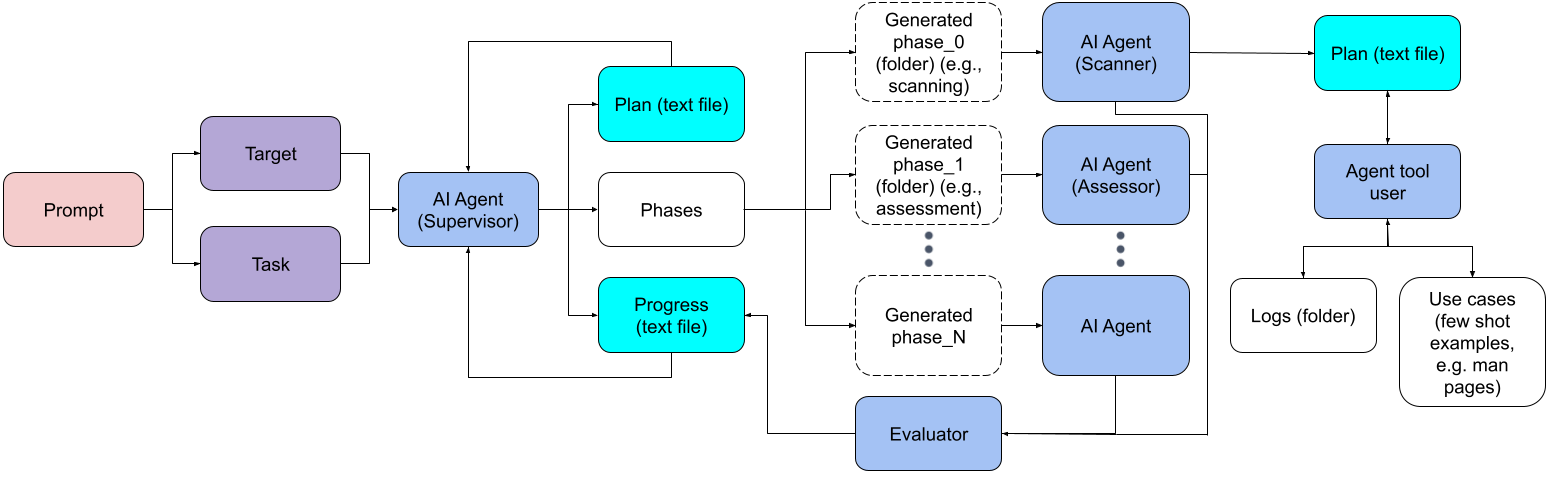}
    \caption{The general workflow of such models}
    \label{Figure 1}
\end{figure*}

\section{Literature Review}
Recent advancements in large language models (LLMs) have significantly impacted the field of cybersecurity, particularly in the automation of penetration testing. Traditionally, penetration testing has been a manual and labor-intensive process, requiring significant expertise and time. However, the introduction of tools like PentestGPT marks a turning point in how these tasks can be automated. PentestGPT leverages the extensive knowledge embedded in LLMs to perform tasks traditionally handled by human penetration testers. This tool has been evaluated using a benchmark created from popular platforms like HackTheBox and VulnHub, which includes 182 sub-tasks aligned with OWASP’s top 10 vulnerabilities. The results indicate a remarkable improvement in task completion rates, with PentestGPT outperforming previous models like GPT-3.5 and GPT-4 by significant margins. This underscores its effectiveness in maintaining context throughout complex testing scenarios, a critical challenge in the application of LLMs to penetration testing tasks \cite{deng-2023}.

In a broader context, the use of generative AI in penetration testing offers both opportunities and challenges. On one hand, generative models can quickly identify vulnerabilities and generate test scenarios that might be missed by human testers. For example, tools like Mayhem utilize techniques such as fuzzing and symbolic execution to uncover vulnerabilities in a fraction of the time it would take a human tester. These models also bring a level of creativity to the process, simulating novel attack vectors that enhance the robustness of penetration testing. On the other hand, challenges remain, particularly regarding the models' ability to fully grasp the broader context of testing scenarios. This can lead to incomplete or inaccurate results, highlighting the need for further refinement of these models to ensure they meet the specific needs of different organizations \cite{hilario-2024}. BreachSeek addresses some of these challenges by employing multiple AI agents to manage context windows, ensuring a more comprehensive understanding throughout the penetration testing process. Unlike other tools, BreachSeek doesn't just generate text-based outputs but also executes commands within a terminal, directly interacting with the target environment.

LLMs are not only transforming penetration testing but are also being integrated into various aspects of cybersecurity. Their applications extend to defensive measures, such as risk management and automated vulnerability fixing. In these areas, LLMs help automate complex tasks, reducing the need for human intervention and allowing for faster, more efficient responses to security threats. However, the effectiveness of LLMs is often limited by their ability to maintain context over extended interactions, a challenge that continues to be a focal point in ongoing research. Future advancements are expected to improve the adaptability of LLMs to specific organizational environments, enabling them to continuously learn and remain effective against evolving cybersecurity threats \cite{motlagh-2024}. Additionally, BreachSeek uniquely contributes to this space by generating a comprehensive, formatted PDF report that captures the entire journey of the penetration testing process, providing valuable insights that are automatically documented and ready for review.

The integration of LLMs into cybersecurity, particularly in automated penetration testing, represents a significant step forward in enhancing security measures. However, these advancements come with their own set of challenges that researchers and practitioners must continue to address to fully realize the potential of these technologies. The continued refinement of tools like PentestGPT, alongside broader applications of generative AI in cybersecurity, will likely shape the future of how organizations defend against increasingly sophisticated cyber threats.

\section{Model Architecture and Implementation}
\subsection{Graph-Based Approach Using LangGraph}
Our model employs a graph-based architecture implemented using LangGraph, enabling the creation of multiple specialized nodes that communicate with each other. This distributed approach offers several advantages:

\begin{enumerate}
    \item Enhanced performance through task distribution across multiple nodes/agents
    \item Flexibility in customizing logic for individual nodes
    \item Mitigation of context window limitations by distributing tasks
\end{enumerate}

\subsection{General Model Workflow}
The general workflow of our model, as illustrated in Figure 1, consists of the following components:

\begin{itemize}
    \item Supervisor: Oversees the entire process, generating action plans and identifying subsequent steps
    \item Specialized agents: Execute specific tasks within their domains of expertise
    \item Evaluator: Assesses the output quality and task completion accuracy
\end{itemize} 

\subsection{Specific Architecture for Penetration Testing}
For this study, we implemented a specialized architecture (Figure 2) that adheres to the general workflow while incorporating task-specific agents:

\begin{enumerate}
    \item Recorder: Maintains a summary of actions and generates a final report when prompted
    \item Pentester: Accesses tools including a shell tool and a Python tool, enabling the utilization of popular penetration utilities in a Kali Linux environment. Its primary role is to execute commands generated by the supervisor and report the output to the evaluator.
\end{enumerate}

\subsection{Implementation Environment}
The model was deployed in a Docker-based Kali Linux environment hosted on RunPod. Key implementation details include:

\begin{itemize}
    \item Development phase: Utilized Anthropic's Claude 3.5 Sonnet model
    \item Testing and future deployment: Plans to use Llama 3.1, an open-source model allowing for customized fine-tuning
\end{itemize}

This architecture and implementation approach allow for a flexible, scalable, and efficient system for automated penetration testing using large language models. The combination of specialized agents, a robust evaluation mechanism, and a supervisory component enables complex, multi-step operations while maintaining coherence and goal-directedness throughout the penetration testing process.

\subsection{Testing Methodology}
For evaluation purposes, a Metasploitable 2 machine was hosted on the same local network as the model. The model was then tasked with exploiting vulnerabilities on this machine, providing a realistic scenario for assessing its penetration testing capabilities.

\begin{figure}[ht]
    \centering
    \includegraphics[width=1\linewidth]{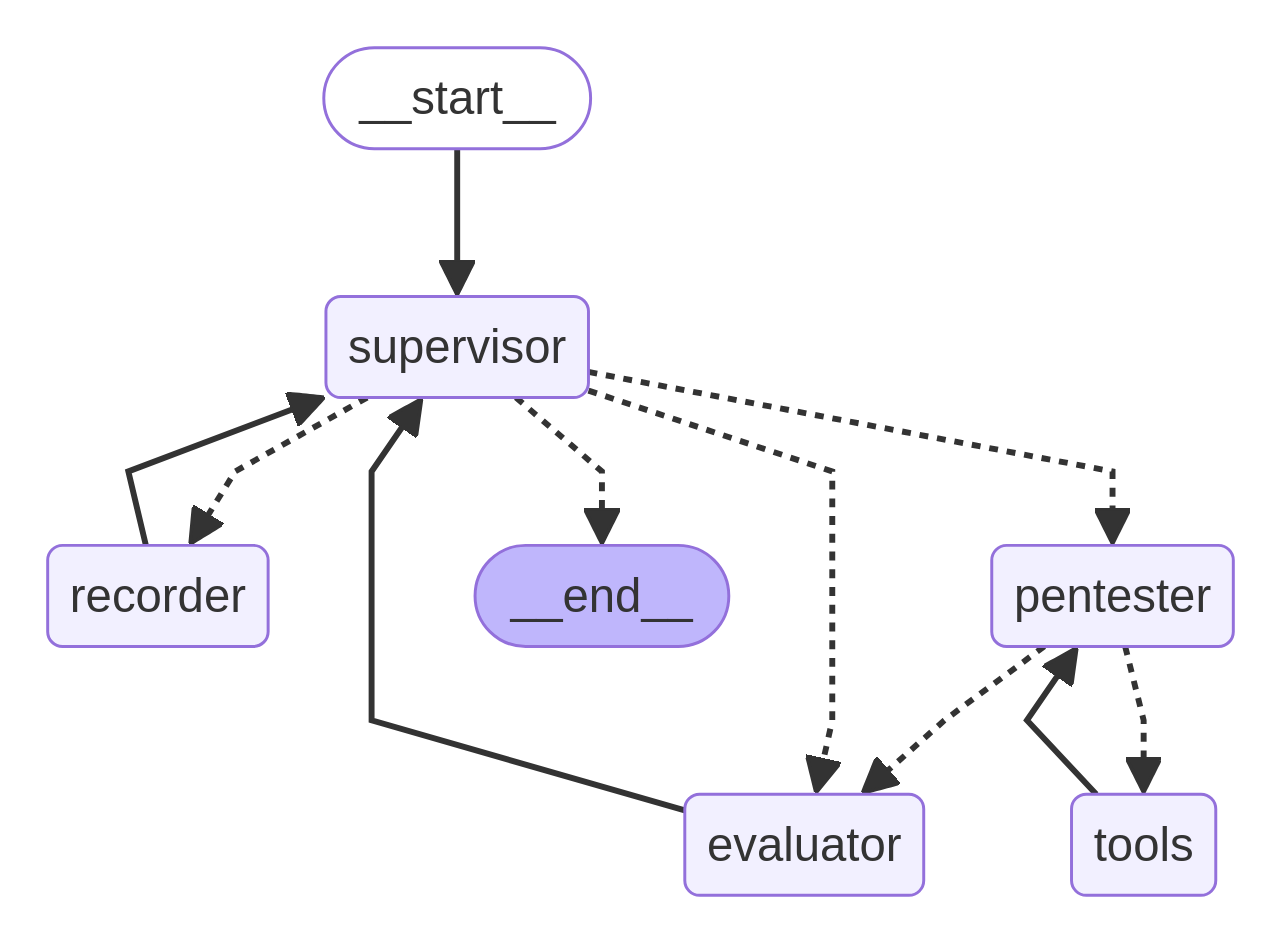}
    \caption{The specific workflow used by our model}
    \label{Figure 2}
\end{figure}

\vspace{-1em}  

\subsection{Web UI}
As part of the product suite we offer, a web UI was developed using NextJS for the front-end and FastAPI for the back-end. A sample from the web UI can be seen in the appendix.

\section{Results}
The efficacy of our model was initially evaluated through qualitative assessment. Future work will incorporate quantitative measures using established benchmarks and standardized examinations. 

Potential benchmarks may include the OWASP Web Security Testing Guide (WSTG) \cite{owasp-2024}. Additionally, we plan to utilize the Offensive Security Certified Professional (OSCP) \cite{offsec-2024} exam content as a standardized measure of performance.

In our preliminary testing, the model successfully exploited a Metasploitable 2 machine, achieving root access with approximately 150,000 tokens. This demonstrates the model's capability to perform complex penetration testing tasks autonomously.

Moreover, our findings suggest that minor adjustments to the workflow and system prompts enable the creation of systems capable of addressing challenges in diverse domains. This versatility indicates the potential for developing general-purpose workflows based on our approach.

\section{Future Work}
\subsection{Human Intervention}
To enhance the safety and control of BreachSeek during penetration testing, future work will focus on integrating a user permission system that prompts for approval before executing specific tools or commands. This feature will allow users to maintain oversight and intervene when necessary, ensuring that critical actions are only performed with explicit consent. This approach not only increases the security of the testing process but also provides a safeguard against unintended or potentially harmful operations.

\subsection{Fine Tuning}
Further development of BreachSeek will involve fine-tuning the model using specialized cybersecurity data. By implementing web scraping techniques to gather cybersecurity write-ups and detailed penetration testing reports, BreachSeek can be trained on a vast array of real-world scenarios and methodologies. This training will enable the system to become more adept at identifying vulnerabilities and recommending effective testing strategies, ultimately improving its performance and reliability in diverse environments.

\subsection{Retrieval-Augmented Generation (RAG)}
To enhance the decision-making process during penetration testing, BreachSeek will incorporate a Retrieval-Augmented Generation (RAG) system. This approach will allow BreachSeek to reference a vector database containing useful penetration testing techniques, strategies, and past experiences. By accessing this database, the system can provide more informed and contextually relevant recommendations, thereby increasing the effectiveness of the testing process.


\subsection{Dynamic and Engaging Responses for Enhanced Interaction}
In future versions of BreachSeek, a key enhancement will be the introduction of dynamic response modes that cater to a variety of user preferences. Instead of restricting the system to purely security-related prompts, BreachSeek will offer users the ability to engage with the model in different styles, including fun and relaxed modes. This flexibility will not only make interactions more enjoyable but also allow users to choose their preferred character or style for chatting.

For those who prefer a more focused approach, BreachSeek will also include a mode that prioritizes task-related communication, minimizing any off-topic chatter. This option ensures that users who need to stay concentrated on security testing can do so without distractions, with the model only responding when the conversation is directly related to the task at hand.

Whether a user prefers a witty companion, a laid-back conversationalist, or a task-focused professional, BreachSeek will adapt to meet these preferences while still providing top-notch security testing services. By expanding the scope of responses and introducing a range of interaction styles, BreachSeek will maintain both its relevance to security tasks and its appeal to a broader audience.

\subsection{Multi-Modality}
To further expand the capabilities of BreachSeek, future iterations will introduce multi-modal input support, allowing users to submit images and videos as part of the penetration testing process. This feature will enable the system to analyze visual content, such as screenshots of network setups or video recordings of security camera feeds, providing a more comprehensive analysis and enabling more sophisticated testing scenarios. By incorporating multiple data types, BreachSeek will be better equipped to handle a broader range of penetration testing challenges.

\section{Conclusion}
BreachSeek, a multi-agent automated penetration testing platform, addresses critical gaps in traditional cybersecurity practices by leveraging Large Language Models through LangGraph. Its graph-based architecture, comprising specialized agents like the supervisor, pentester, and recorder, enables efficient task distribution and mitigates context window limitations. Deployed in a Docker-based Kali Linux environment, BreachSeek demonstrated its effectiveness by successfully exploiting a Metasploitable 2 machine within 150,000 tokens. While initially evaluated qualitatively, future work will incorporate quantitative measures using benchmarks like OWASP WSTG and OSCP exam content. Planned enhancements include a user permission system for human oversight, fine-tuning with specialized cybersecurity data, integration of Retrieval-Augmented Generation (RAG), enhanced dynamic and responsive interactions according to user preference, and multi-modal input support. These advancements, coupled with BreachSeek's ability to generate comprehensive security reports, position it as a powerful, adaptable tool in the evolving landscape of AI-driven cybersecurity solutions, promising continued innovation in automated penetration testing and defense against sophisticated cyber threats.

The code used for the model can be found here: \url{https://github.com/snow10100/pena/}

\printbibliography

\newpage
\onecolumn  
\appendix
\section{Appendix}
\begin{figure}[ht]
    \centering
    \includegraphics[width=0.9\textwidth]{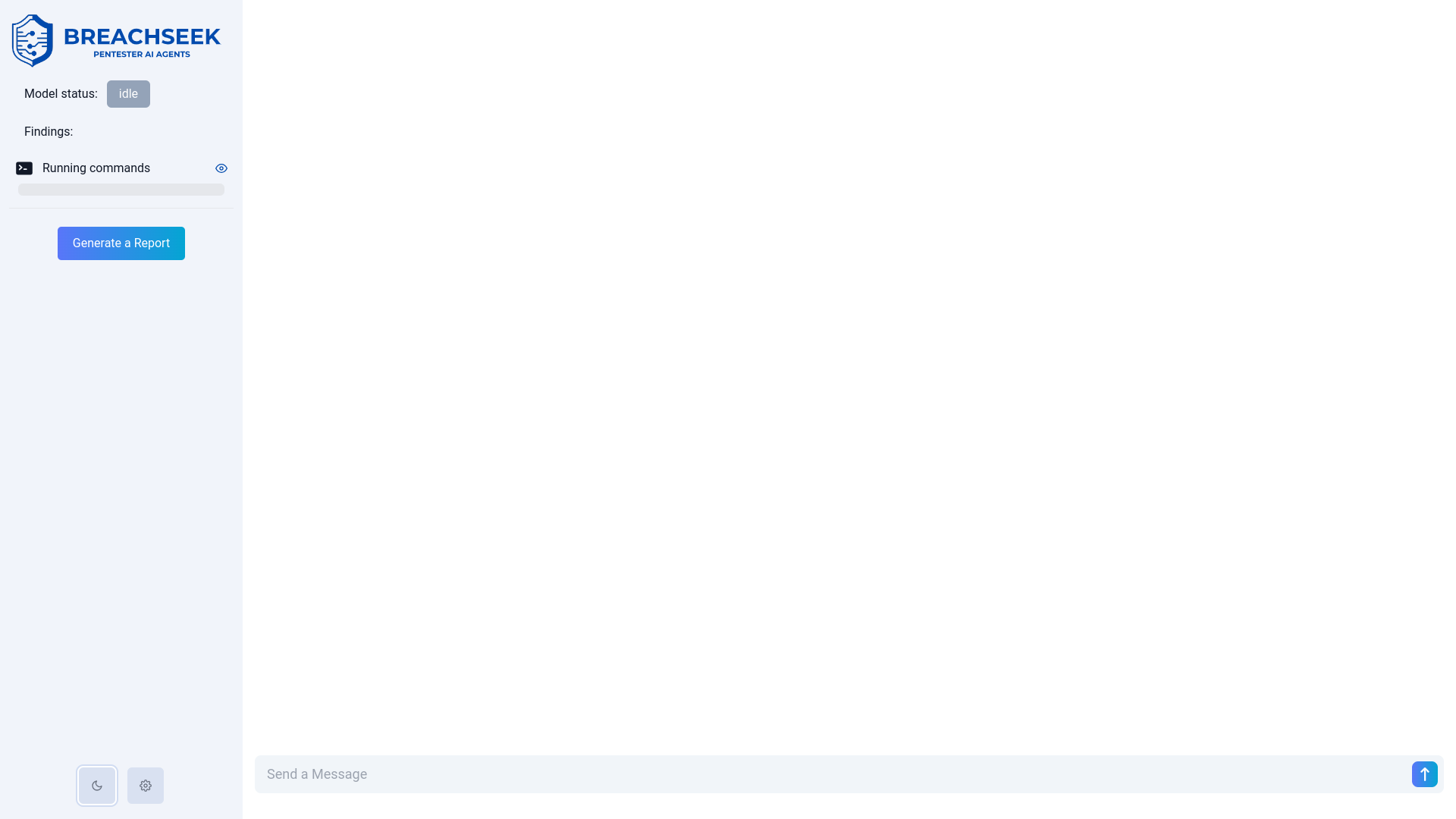}
    \caption{The clean web UI when you start chatting with model}
    \label{Figure 3}
\end{figure}

\begin{figure}[ht]
    \centering
    \includegraphics[width=0.9\textwidth]{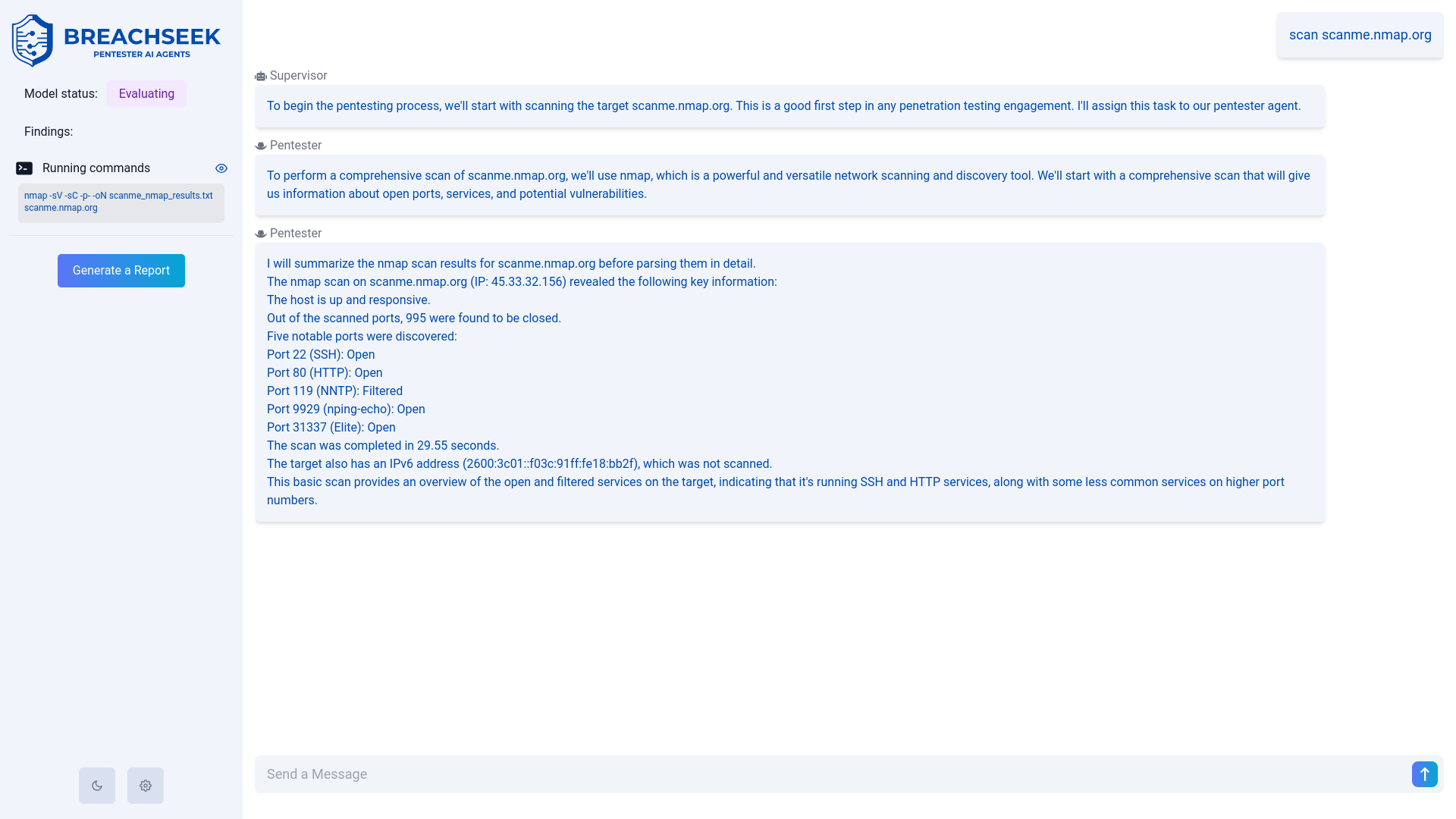}
    \caption{The AI agents performing a task}
    \label{Figure 4}
\end{figure}

\begin{figure}[ht]
    \centering
    \includegraphics[width=0.9\textwidth]{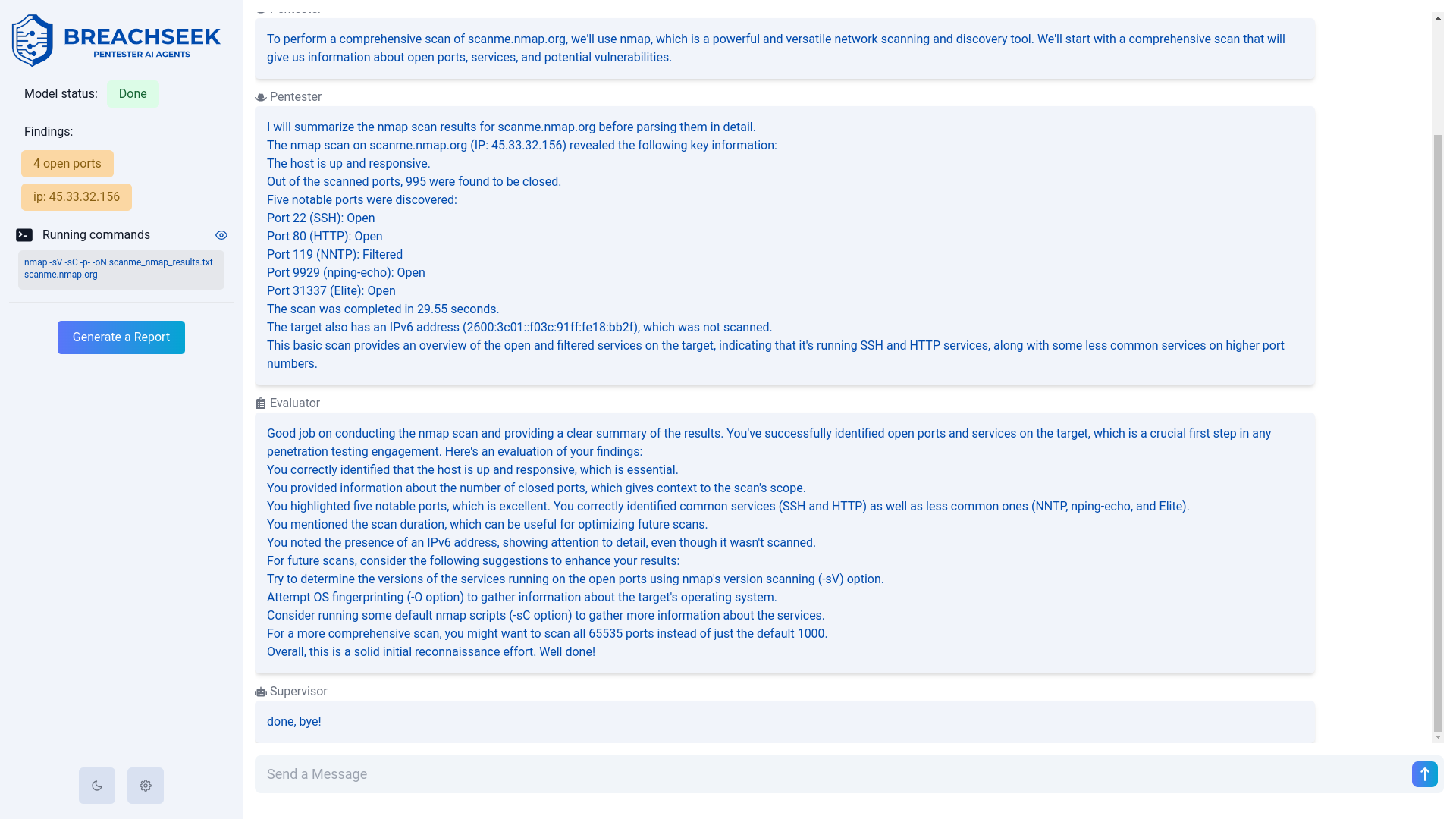}
    \caption{The web UI when the task is done}
    \label{Figure 5}
\end{figure}

\begin{figure}[ht]
    \centering
    \includegraphics[width=0.9\textwidth]{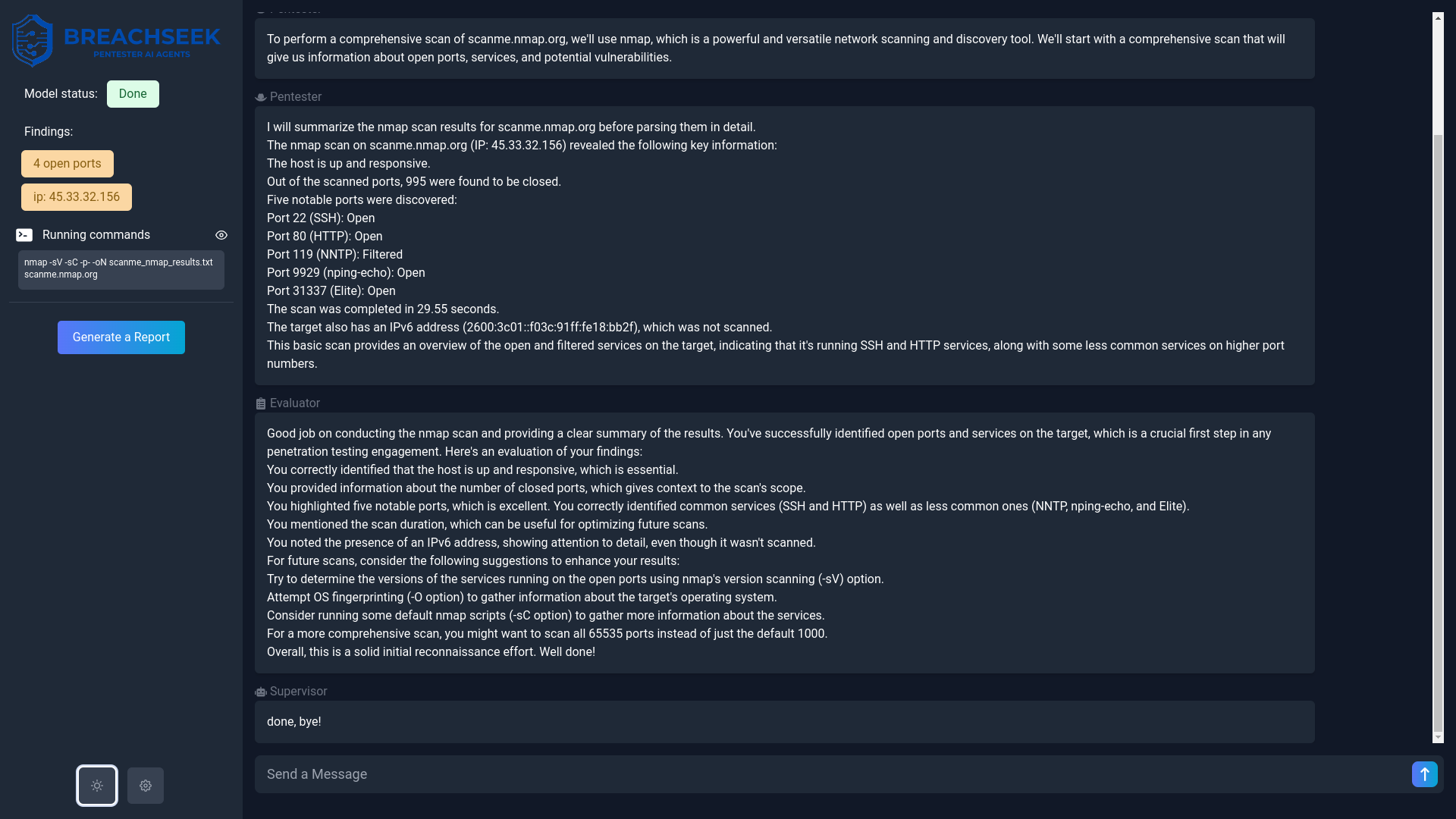}
    \caption{Web UI Dark mode}
    \label{Figure 6}
\end{figure}

\end{document}